# Epitaxial Synthesis of Blue Phosphorene

*Wei Zhang, Hanna Enriquez, Yongfeng Tong, Azzedine Bendounan, Abdelkader Kara, Ari P. Seitsonen, Andrew J. Mayne, Gérald Dujardin, and Hamid Oughaddou\**

W. Zhang, Dr. H. Enriquez, Dr. A. J. Mayne, Dr. G. Dujardin, Prof. H. Oughaddou
Institut des Sciences Moléculaires d'Orsay, ISMO-CNRS, Bât. 520, Université Paris-Sud, F-91405 Orsay, France
E-mail: Hamid.oughaddou@u-psud.fr
Dr. Y. F. Tong, Dr. A. Bendounan
TEMPO Beamline, Synchrotron SOLEIL, L'Orme des Merisiers Saint-Aubin, B.P.48, F-91192 Gif-sur-Yvette Cedex, France
Prof. A. Kara
Department of Physics, University of Central Florida, Orlando, FL 32816, USA
Dr. A. P. Seitsonen
Département de Chimie, Ecole Normale Supérieure, F-75231 Paris Cedex 05, France
Prof. H. Oughaddou
Département de physique, Université de Cergy-Pontoise, F-95031 Cergy-Pontoise Cedex, France

**Abstract:**

Phosphorene is a new two-dimensional material composed of a single or few atomic layers of black phosphorus. Phosphorene has both an intrinsic tunable direct band gap and high carrier mobility values, which make it suitable for a large variety of optical and electronic devices. However, the synthesis of single-layer phosphorene is a major challenge. The standard procedure to obtain phosphorene is by exfoliation. More recently, the epitaxial growth of single-layer phosphorene on Au(111) has been investigated by molecular beam epitaxy and the obtained structure has been described as a blue-phosphorene sheet. In the present study, large areas of high-quality monolayer phosphorene, with a band gap value at least equal to 0.8 eV, have been synthesized on Au(111). Our experimental investigations, coupled with DFT calculations, give evidence of two distinct phases of blue phosphorene on Au(111), instead of one as previously reported, and their atomic structures have been determined.

Even though graphene has a very high potential in electronics, its semi-metallic character[1] is a significant limitation for a large variety of electron devices. This limitation can be overcome by the use of new emerging two-dimensional (2D)materials such as silicene,[2-5] germanene,[6-8] transition metal di-chalcogenides,[9-11] hBN,[12-14] and phosphorene.[15-17] Phosphorene is composed of a single layer of black phosphorus, presents electronic and optical properties that are ideally suited to future electronic/optoelectronic devices because of the combination of an intrinsic tunable direct band gap, with high carrier mobility.[18] In addition phosphorene has a special property: stacking phosphorene layers together enables the band-gap to be shifted significantly from the infra-red to the visible range, since bulk black phosphorus[19-21] has a gap 0.35 eV, and one monolayer of phosphorene[22] a gap of 1.5 eV.

However, up to now the synthesis of phosphorene has been achieved mainly using exfoliation of bulk black phosphorus. Only basic electronic prototype devices have been reported, preventing any implementation into large scale electronic circuits. The demonstration of molecular beam epitaxy (MBE) growth of blue phosphorene, a new allotrope of black phosphorene, has been reported only very recently.[23-25] Indeed, since phosphorene is a very promising novel material for optical and electronic technology, the controlled synthesis of phosphorene, based on industrial processes such as MBE and chemical vapor deposition (CVD), is a necessary step for large scale applications.

Different allotropes of phosphorus exist of which the most stable is black phosphorus,[26] presenting a layered structure similar to graphite. However, unlike graphene, which is planar, phosphorene exhibits a strongly puckered structure.[27] **Figure 1a** shows a black phosphorene layer characterized by armchair ridges when viewed from the side. It has been reported[23] that a specific dislocation in black phosphorene induces a new structure with zigzag puckering, called blue phosphorene (Figure 1b). The lattice constants are $a_1 = 0.437$ nm and $a_2 = 0.331$ nm for black phosphorene,[20,23,24] and $a_1 = a_2 = 0.328$ nm for blue phosphorene.[24,28]

Up to now, only a few studies have reported the growth of blue phosphorene (bl-P hereafter) on a Au(111) substrate using the MBE method.[23-25] There it has been shown that deposition of 1 monolayer (ML) of phosphorus induces a (5 × 5) periodic structure. However, the proposed model of the atomic structure is under debate. Indeed, a number of discrepancies exist between the published scanning tunneling microscope (STM) images and the calculated structures reported in these articles.

In this article, we present a detailed investigation of the growth of blue phosphorene on Au(111) using MBE. Several surface sensitive techniques are employed: low energy electrons diffraction (LEED), *in situ* low temperature scanning tunneling microscopy (LT-STM), photoemission spectroscopy (PES), and angle resolved photoemission spectroscopy (ARPES). The epitaxy of blue phosphorene on Au(111) induces two different structures exhibiting both the (5 × 5) periodicity in contrast with the one structure reported in ref. 23. The goal of this work is to focus mainly on a new extended structure of blue phosphorene exhibiting three protrusions in the STM images instead of six.

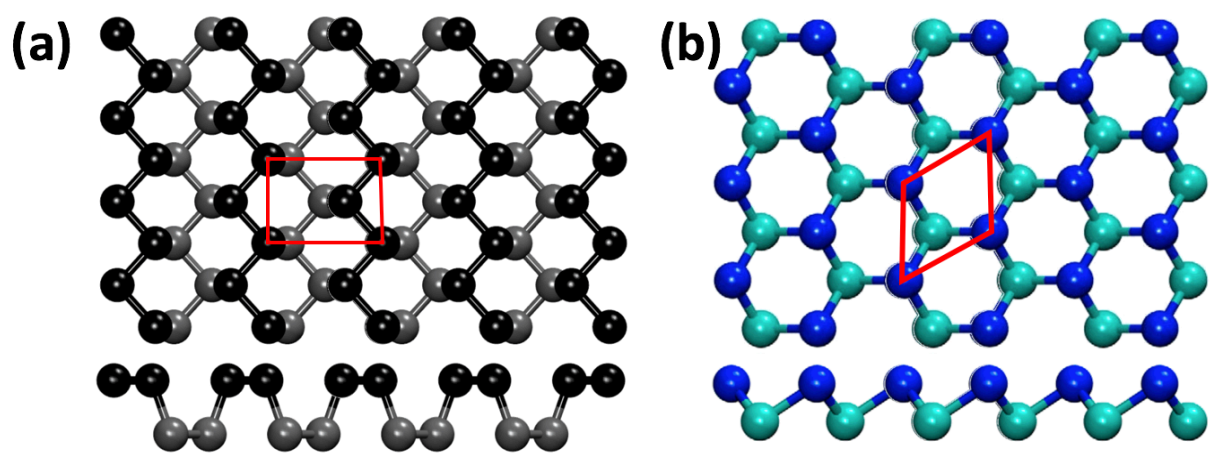

**Figure 1.** Top and side view of black (a) and blue (b) phosphorene. Atoms in different planes are indicated with different colors (top darker/blue, bottom lighter/cyan). The unit cell of each structure is indicated in red.

**Figures 2a and 2b** show the LEED patterns recorded at 67 eV, corresponding to the clean Au substrate and 1 ML of phosphorus on the Au(111) surface after deposition, respectively. The Au(111) surface was held at 260°C during the deposition. In Figure 2a we observe the (1 × 1) spots surrounded by satellites corresponding to the well-known (22 ×√3) reconstruction of Au.[5] In Figure 2b, the deposition of phosphorene yields a (5 × 5) structure in agreement with one reported in refs. 23-25. For greater clarity, the six outer symmetric spots originating from the Au(111) substrate are indicated by full circles in both images. We also observe in Figure 2b that the $\frac{4}{5}$ spots along the main crystallographic directions of Au(111) are much brighter than the other spots. They are highlighted by dashed circles.

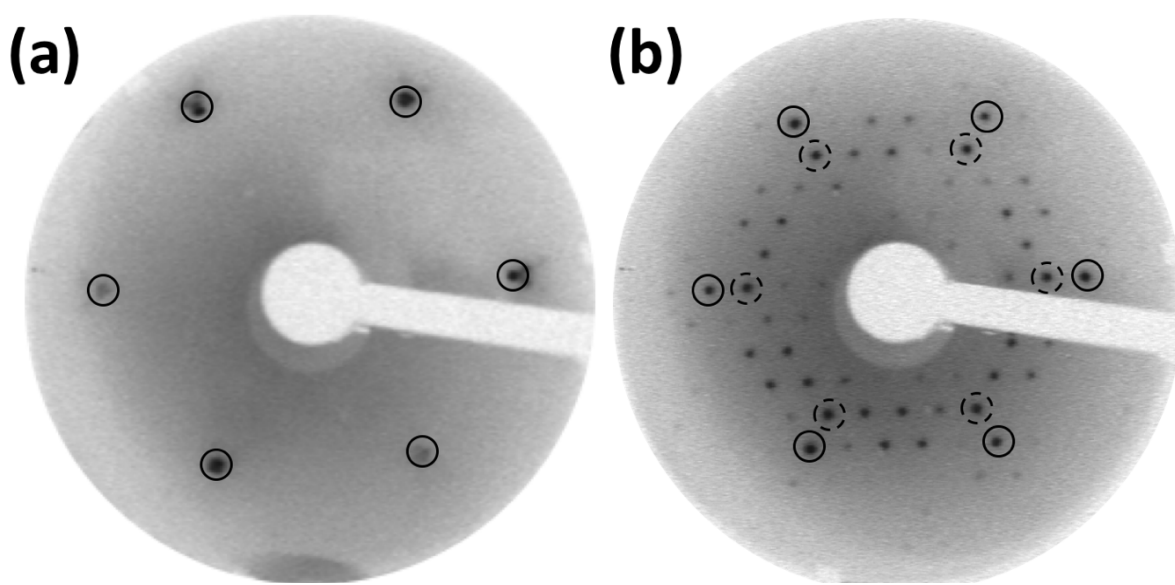

**Figure 2.** LEED pattern recorded at 67 eV for the clean Au substrate (a) and after deposition of 1 ML of phosphorus on Au(111) surface held at 260°C (b). The (1 ×1) spots of Au are highlighted by full circles in both images while in (b), the $\frac{4}{5}$ spots along the main crystallographic directions of Au(111) are highlighted by dashed circles.

**Figure 3** shows a STM image of the first stage of growth of phosphorus on Au(111) kept at 260°C during the deposition. We observe a periodic lattice composed of triangles. Two sizes of triangles coexist composed of three or six protrusions as indicate. Both structures are consistent with the (5 × 5) periodicity observed in LEED pattern. Such local structures were also reported in ref. 23.

However, in our investigation, we observed only a few small areas formed by triangles with six protrusions within the long range ordered structure. We found that the most prevalent structure of the first deposited layer is composed of triangles with three protrusions, which cover large areas of the substrate, as shown in **Figure 4**. This is in contrast with the studies reported in refs. 23 and 25 where it was concluded that the phosphorene structure obtained on Au(111) was composed only of triangles containing six bright protrusions.

This can be explained by the different growth conditions, in particular the temperature of the substrate. In addition, the authors in ref. 24 used InP as source to deposit phosphorus instead of molecular beam epitaxy in our case and those of refs. 23 & 25. It is clear that the decomposition of InP may produce different phosphorus precursors that will yield different structures, and that the temperature of the sample during the growth is a key parameter. Indeed, we would expect phosphorus atoms could diffuse in the bulk at higher temperatures.

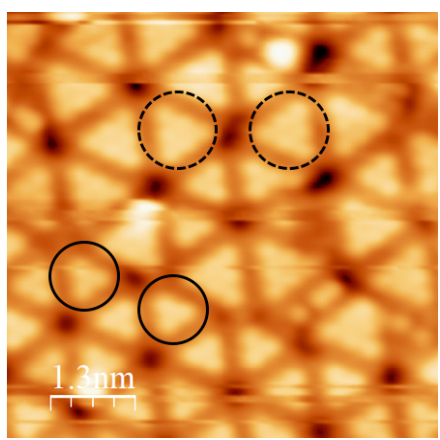

**Figure 3.** Filled state STM image corresponding to 1 ML of phosphorus on Au(111) held at 260°C (U=− 2.0 V, I=0.44 nA). Two triangles with 3 and 6 protrusions each are highlighted by full and dashed circles, respectively.

A typical STM topography recorded after deposition of 1 ML of phosphorene on Au(111) is shown in Figure 4a. Phosphorus covers the surface terraces with a periodic hexagonal structure. Figure 4b shows a magnified STM image of Figure 4a where we observe clearly an array of bright trimers, with two phosphorus trimers per unit cell. Figure 4c shows the line profile measured along line A in Figure 4b. This line profile shows a periodicity of 1.44 nm which corresponds to five times the nearest-neighbor distance of Au(111) surface (5 × 0.288 nm = 1.44 nm) in good agreement with the (5 × 5) structure shown in the LEED experiments. This line profile also gives a distance between the two nearest-neighbor bright protrusions of 0.35 nm which is larger than the calculated phosphorene nearest-neighbor atom distance of 0.222 nm.[20] However, this distance is close to the lattice constant of blue phosphorene, where it corresponds to the second neighbor distance of phosphorus atoms.[23] The line profile measured along line B gives a distance of 0.69 nm between two neighboring triangles.

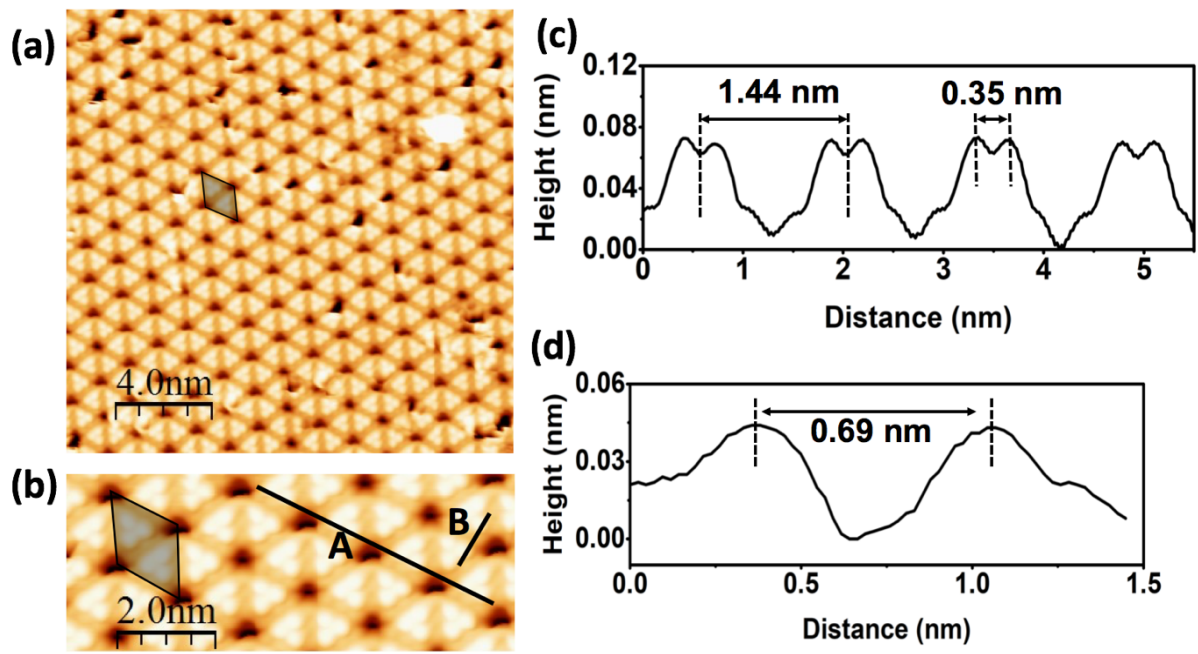

**Figure 4.** (a) Filled state STM image of 1 ML of phosphorus deposited on Au(111) kept at 260°C, the unit cell is shown (U =−250 mV, I=1.34 nA). (b) Zoom of (a) showing that a unit

cell includes two trimers of phosphorene atoms (U=−100 mV, I=2.60 nA). (c) Profile along line A in image (b). (d) Profile along line B in image (b).

In the LEED pattern shown in Figure 2b, we observe that the $\frac{4}{5}$ spots along the main directions of Au(111) are much brighter than the other spots. This diffraction spot arises from two contributions: the diffraction of the (5 × 5) superstructure superposed on the diffraction of the (1 × 1) sheet made by the phosphorus atoms. This indicates that 5 times the nearest-neighbor distance of Au(111) (5 ×0.288 nm = 1.44 nm) is equal to 4 times the lattice constant of phosphorus (4 × 0.36 nm = 1.44 nm) in the phosphorus sheet. This value of 0.36 nm deduced for the phosphorus is very close to the one extracted from the STM images (0.35 nm), which is only 0.02 nm larger than the expected distance from standalone bl-P (0.328 nm).[23] This difference can be explained by the effect of the substrate on the sheet of phosphorene. The observed distance between the neighboring protrusions in the STM image (Figure 4b) corresponds to the distance between second nearest-neighbor phosphorus atoms. This suggests that the (5 × 5) sheet is formed by the honeycomb structure of the bl-P sheet. We can then deduce a projected distance of the nearest-neighbor distance of the phosphorus atoms of 0.202 nm. Combining the results of the LEED and the high-resolution STM images, we deduced that the deposition of 1 ML of phosphorus on Au(111) gives rise to a phosphorene layer with the bl-P structure.

As the structural parameters of bl-P are different from those of Au, it is not possible to adsorb a commensurate extended film on the Au(111)-(5×5), which has a lattice parameter of 14.77 Å. The lattice parameter for free-standing bl-P was found to be 0.3275 nm from DFT, yielding lattice parameters of 0.1310 and 0.1638 nm for bl-P (4 × 4) and (5 × 5), respectively. The Au (5 × 5) lattice parameter falls right in between those for the bl-P (4 × 4) and (5 × 5). Squeezing a bl-P (5 × 5) in the Au (5 × 5) would induce a huge strain in the system, which would lead to a large unfavorable energy. On the other hand, fitting a bl-P (4 × 4) on the Au (5 × 5) resulted in a stretched bl-P sheet with two neighboring bl-P 16-atom triangles. The adsorbed triangles were placed to preserve the high symmetry of the substrate as can be seen in the top part of **Figure 5a**. After full relaxation of the system, the P-P distance along the sides of the triangles was found to be around 0.34 nm, close to the lattice parameter of the bl-P, however, the corner-to-corner distance between triangles in neighboring unit cells was found to be around 0.44 nm, introducing a slight discontinuity of 0.1 nm, in the otherwise extended bl-P. In these two triangles of 16 atoms each, we found that 6 atoms adsorb at a height of about 0.22 nm above the remaining 10 atoms. These 2 × 6 atoms are the ones observed in both the experimental (Figure 3) and the simulated STM images (Figure 5a-bottom).

The model presented in Figure 5-a is close to the proposed model in Ref. 25. However, it cannot explain the experimental STM image shown in Figure 4a, where triangles with three protrusions are observed. According to the same study, no more than one additional phosphorus atom could be adsorbed per unit cell on the first-layer phosphorene. However, our experimental STM topography (Figure 4a) can be only explained by adding 2 × 3 extra P atoms per unit cell on top of the first phosphorene layer in agreement with the calculations discussed below.

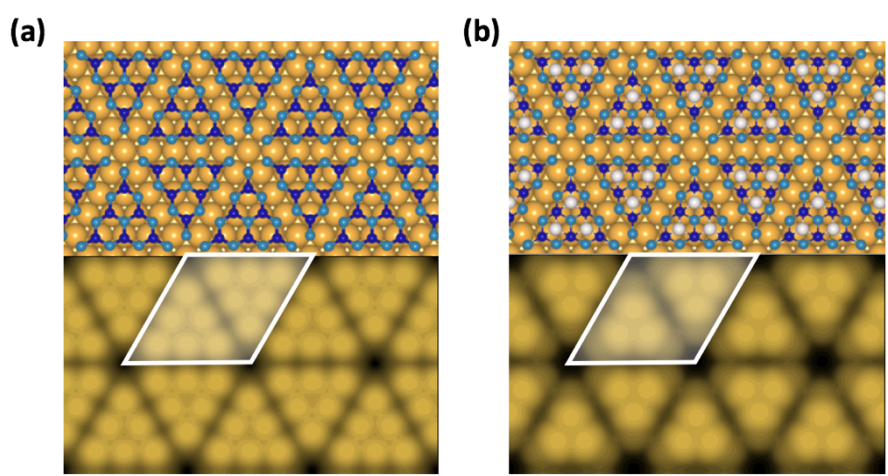

**Figure 5.** The model systems of blue phosphorene on the (5 × 5) Au(111): (a)-top 32 P atoms forming two triangles with 6 bright protrusions each; (a)-bottom the corresponding simulated STM image. (b)-top additional 2 × 3 P atoms on the two triangles; (b)-bottom the corresponding STM image.

The model of this new super-structure is shown in Figure 5b-top. It consists of 38 P atoms with 3 atoms adsorbed on each of the triangles in the first model, which gives a bl-P second layer via A-B-C stacking. The fully relaxed system presents a P-P distance of 0.33 nm, within a 3-atom triangle, and a lateral distance of 0.663 nm between the two 3-atom triangles. Both these distances are in good agreement with the distances observed in the experimental images of 0.35 and 0.69 nm (see Figures 4c and d). In this extra 2 × 3 phosphorus atom system, the 3 atoms on top of each 16-atom triangles again lie at about 0.22 nm above the 6 top atoms of the first phosphorene layer and correspond to the spots observed in experimental (Figure 4b) and calculated STM images (Figure 5b-bottom).

We have also performed photoemission spectroscopy measurements to characterize this structure further. The $P_{2p}$ and $Au_{4f}$ core level spectra recorded at normal emission are shown in **Figure 6**. All spectra are fitted with a Doniach-Sunjic line shape.[29] The $P_{2p}$ spectrum (Figure 6a) is reproduced with two spin-orbit split components S1 and S2 located respectively at 128.88 and 129.17 eV. The best fit was obtained with a 166 meV Gaussian profile and an 80 meV Lorentzian profile, while the spin-orbit splitting is 0.865 meV and the branching ratio is 0.45. The presence of two components indicates that the phosphorene atoms have only two chemical environments. The S1 component is assigned to the phosphorus atoms with only a phosphorus environment while the S2 component is assigned to phosphorus atoms with a gold environment, in good agreement with the model derived from the DFT calculations. Indeed, the models present two phosphorene environments; P-P and P-Au.

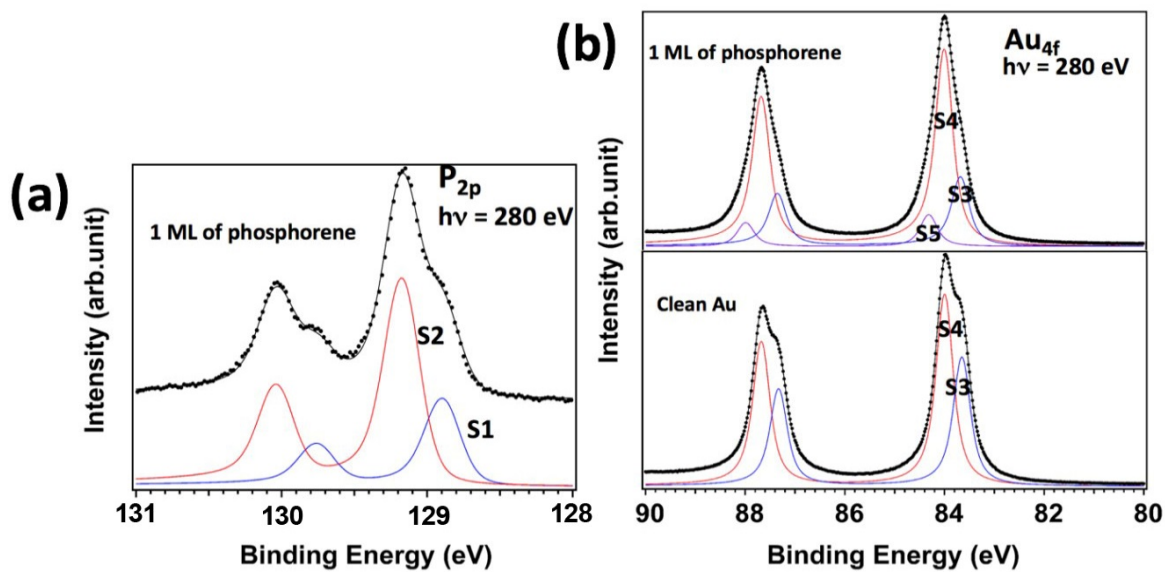

**Figure 6.** P$_{2p}$ and Au$_{4f}$ core levels spectra (dots) and their de-convolutions (solid line overlapping the data points) recorded at hν = 280 eV in normal emission. (a) The P$_{2p}$ spectrum is reproduced with two components as well as the Au$_{4f}$ corresponding to the clean Au(111) (b-lower panel). After deposition, Au$_{4f}$ spectrum is reproduced with three components (b-upper panel).

The Au$_{4f}$ spectrum recorded for the clean gold (Figure 6b) is also reproduced with two spin-orbit split components S3 and S4 located respectively at 83.65 and 84.00 eV. The S3 and S4 components are attributed to the surface and bulk environment of Au, respectively. The best fit was obtained with a 133 meV Gaussian profile and a 320 meV Lorentzian profile, while the spin-orbit splitting is 3.678 eV and the branching ratio is 0.75.

Figure 6b also shows the Au$_{4f}$ spectrum recorded after deposition of 1 ML of phosphorene. The fit was obtained with the same parameters as for the clean Au surface. Here, the spectrum is again reproduced with the two spin-orbit split components (S3 and S4) as for the clean gold. However, it was necessary to add an additional component (S5) located at 84.32 eV that we attribute to the gold-phosphorus environment. We observe that the S3 component attributed to the surface environment of Au does not vanish completely, indicating that a few small areas of the bare surface remain, even though none was found in the many STM images. This may be due to some bare areas of the Au surface remaining. Previous studies have reported X-ray photoemission spectroscopy (XPS) measurements.[23] However, they were not able to provide a quantitative analysis of the spectra because they had a conventional laboratory source, which did not give a high resolution as we have obtained with Synchrotron radiation.

We have also studied the valence band of phosphorene on Au(111) with ARPES. Here we probe the band structure of the new phosphorene structure composed of trimer protrusions (see Figure 4a and 5b). **Figure 7a** shows ARPES results recorded at room temperature (RT) at 66 eV photon energy, along the $\Gamma - K$ and the $\Gamma - M$ directions of the substrate. At this photon energy the intensity of the Au-sp bands is the most intense; hence, even after the P deposition, the Au-sp bands are still visible.

Between 0 and 0.8 eV, we observe only the filled states of Au which gives a relatively weak intensity when covered by the phosphorus monolayer. However, for binding energies of more than 0.8 eV, the intensity is much higher due to the additional contribution of the phosphorene valence band. This indicates that the band gap of phosphorene is larger than 0.8 eV, and its width depends on the position of the conduction band of phosphorene, which we cannot, however, measure with the ARPES. The Au *sp* bands are clearly observed on the left and right of the *K* and *M* points, respectively.

Note that the vertical stripes continuing above 0.8 eV come from the inhomogeneity of the analyzer detector. Figure 7b shows the integrated ARPES signal showing clearly the appearance of the band gap. We note that the ARPES measurements sample a larger area giving unambiguously a band gap of at least 0.8 eV which is consistent with the reported value of 1.18 eV in ref 25.

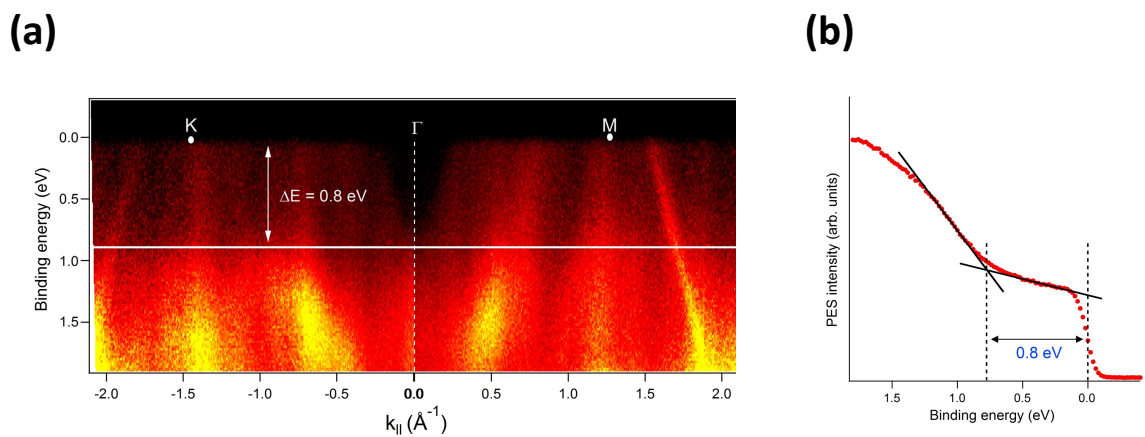

**Figure 7.** (a) ARPES measurements recorded at hv = 66 eV after deposition of monolayer of phosphorene on clean Au(111). The band gap is clearly observed. (b) The integrated signal showing also the existence of the band gap.

In conclusion, we have shown that 2D blue phosphorene can be synthesized on Au(111) using the MBE method. At a coverage close to 1ML, the phosphorus layer forms a (5 × 5) superstructure with a band gap of at least 0.8 eV. Moreover, the results from the STM and LEED measurements coupled with DFT calculations demonstrate that the new blue phosphorene structure consists of a stretched first layer of blue phosphorene, with additional trimers of the second layer at higher coverage. This new model is consistent with both the (5×5) periodicity and the extended periodic structure observed in the experimental STM topography.

**Experimental Section**

The ultra-high vacuum (UHV) apparatus in which the experiments were performed is equipped with the standard tools for surface preparation and characterization: an ion gun for surface cleaning, Low Energy Electron Diffraction (LEED), an Auger Electron Spectrometer (AES) for chemical surface analysis, and an Omicron Scanning Tunneling Microscope (LT-STM) for atomic scale surface characterization at 77 K. The photoemission measurements were performed at the Synchrotron SOLEIL-France on the TEMPO beam-line on the same sample as used in LEED and STM characterization. We used commercial Au(111) crystals with 99,999% purity. The Au(111) substrate was cleaned by several sputtering cycles (650 V Ar$^+$ ions, P = 1× 10$^{-5}$ mbar) followed by annealing at 500°C for 40 minutes.

A Knudsen source loaded with black phosphorus was used to grow phosphorene on the Au(111) substrate. The Au(111) surface was held at 260°C during the deposition. During the phosphorus deposition, the surface structure, thickness, and the chemical structure were controlled by AES and LEED systematically. All STM experiments were carried out at 77 K and the base pressure during the experiments was < 3.0 × 10$^{-10}$ mbar.

To support the experimental conclusions, we performed DFT calculations using the projector augmented wave (PAW) method[30,31] and the Perdew-Burke-Ernzenhof generalised gradient approximation (GGA-PBE)[32] functional as implemented in the Vienna ab initio simulation package (VASP) version 5.4.1.[33] To model the unreconstructed Au(111) surface, we used a 5 layer slab with periodic boundary conditions. Each layer contained 25 atoms, representing the (5 × 5) structure. A 15-17 Å of vacuum space was placed vertically between slabs. We used a plane-wave energy cutoff of 255 eV. The Brillouin zone was sampled using a 4 × 4 × 1 k-point mesh, and the force criterion was set at 0.01 eV/Å. The bottom two layers of the substrate were kept fixed at their bulk positions, while the top three layers and all the phosphorus atoms were allowed to relax. The calculated STM images were produced from the partial charge density of the relaxed systems using the Tersoff-Hamann method.[34]

**Notes**

The authors declare no competing financial interest.

**ACKNOWLEDGEMENTS**

W.Z. would like to thank the Chinese Scholarship Council (CSC) for the Ph.D financial support (scholarship). A.K. acknowledges ISMO-CNRS, Paris-Sud University, partial support from the